\documentclass[aps,prl, reprint,superscriptaddress, preprintnumbers]{revtex4-1}
\usepackage{blindtext}
\usepackage[english]{babel}
\usepackage{amsmath}
\usepackage{wasysym}
\usepackage{graphicx}
\usepackage{caption}

\pdfoutput=1

\def\be{\begin{equation}}
\def\ee{\end{equation}}
\def\bea{\begin{eqnarray}}
\def\eea{\end{eqnarray}}
\def\ben{\begin{equation}}
\def\een{\end{equation}}
\def\bean{\begin{eqnarray}}
\def\eean{\end{eqnarray}}

\def\pd{\partial}
\def\a{\alpha} 
\def\b{\beta}
\def\g{\gamma}
\def\d{\delta}
\def\m{\mu}
\def\n{\nu}

\def\l{\lambda}

\def\r{\rho}

\def\bR{\bar{R}}

\def\bg{\bar{g}}

\def\bp{\bar{\phi}}
\def\s{\sigma}
\def\e{\epsilon}

\def\bma{\begin{pmatrix}}
\def\ema{\end{pmatrix}}

\def\f{\phi}

\def\D{\nabla}

\def\bg{\bar{g}}
\def\bi{\begin{itemize}}
\def\ei{\end{itemize}}
\def\bn{\bar{\nabla}}

\def\bp{\bar{\phi}}

\begin{document}

	\preprint{IFT-UAM/CSIC-15-112}
	\preprint{FTUAM-15-34}
	\preprint{FTI-UCM/161}
	
\title{Conformal  Dilaton Gravity: Classical Noninvariance Gives Rise To Quantum Invariance}
\author{Enrique \'Alvarez}
\email{enrique.alvarez@uam.es}
\affiliation{ Instituto de F\'{\i}sica Te\'orica, IFT-UAM/CSIC, Universidad Aut\'onoma, 28049 Madrid, Spain}
\affiliation{Departamento de F\'{\i}sica Te\'orica, Universidad Aut\'onoma de Madrid, 28049 Madrid, Spain}
\author{Sergio Gonz\'alez-Mart\'{\i}n}
\email{sergio.gonzalez.martin@csic.es}
\affiliation{ Instituto de F\'{\i}sica Te\'orica, IFT-UAM/CSIC, Universidad Aut\'onoma, 28049 Madrid, Spain}
\affiliation{Departamento de F\'{\i}sica Te\'orica, Universidad Aut\'onoma de Madrid, 28049 Madrid, Spain}
\author{Carmelo P. Mart\'{\i}n}
\email{carmelop@fis.ucm.es}
\affiliation{Universidad Complutense de Madrid (UCM), Departamento de F\'isica Te\'orica I,
Facultad de Ciencias F\'{\i}sicas, Av. Complutense S/N (Ciudad Univ.),
28040 Madrid, Spain}

\begin{abstract}
When quantizing Conformal Dilaton Gravity there is a conformal anomaly which starts at two loop order. This anomaly stems from evanescent operators on the divergent parts of the effective action. The general form of the  finite counterterm which is necessary in order to insure cancellation of the Weyl anomaly to every order in perturbation theory has been determined using only  conformal invariance . Those finite counterterms do not have any inverse power of any mass scale in front of them (precisely because of conformal invariance) and then they are not negligible in the low energy deep infrared limit.
The general form of the ensuing modifications to the scalar field equation of motion has been determined and some physical consequences extracted.
\end{abstract}
\maketitle

\section{Introduction}
It is well known \cite{Adler} that an {\em anomalous} term in a Ward identity only qualifies as a true anomaly when there is no local counterterm that can be added to the action in such a way that it cancels the putative anomalous piece of the identity. Put it in another way, there is a consistency condition, the Wess-Zumino consistency condition \cite{Zumino}, which is a reflection on the gauge algebra acting on the effective action. If an anomalous variation of the action under a gauge transformation with parameters $\Lambda^a$ appears
\be
{\d W\over \d \Lambda_a(x)}\equiv G^a(x)
\ee
(Although this formalism has been developed with non-abelian gauge anomalies in mind, it can easily be adapted to conformal anomalies as well \cite{Boulanger}).
Then the consistency conditions read
\be
{\d G^a(x)\over \d \Lambda_b(y)}-{\d G^a(y)\over \d \Lambda_b(x)}= f_{abc}\d(x-y)G_c(x)
\ee
True anomalies are then solutions of the consistency equations which are not themselves variations, that is, that there is no local lagrangian such that
\be
G_a(x)={\d C\over \d \Lambda_a}
\ee	
Defining the contraction of the anomaly with the ghosts
\be
G^1\equiv c^a G_a
\ee
(the superindex as a reminder of the ghost number)
the consistency relationships can be written in a sophisticated way as 
\be
s~ G^1=d \a^2
\ee
The appearance of a total differential on the second member is due to the fact that it is only necessary for it to vanish when integrated.
The demand that the anomaly is not trivial reads in BRST language,
\be
G^1\neq s G^0 + d \b^0
\ee
\par 
In a recent paper  \cite{Alvare} (whose notation will be followed here) we have examined an apparently quasi-trivial theory, namely what we have dubbed {\em conformal dilaton gravity} (CDG). In it the Weyl parameter is upgraded to a St\"uckelberg field to compensate the Weyl transformation of the Einstein-Hilbert lagrangian.
\par
What we have found rests on  the mild assumption that what was true at one loop (namely that the on-shell counterterm is the Weyl transform of the Einstein-Hilbert counterterm) remains true to two loops, so that the two loop counterterm will also be the Weyl transform of the Goroff-Sagnotti \cite {Goroff} one. With this assumption we found a two-loop Weyl anomaly in our {\em trivial} theory. This anomaly is however trivial in the sense that it can be eliminated by a counterterm, which is not strictly local, involving logarithms of the physical fields.
\par
This is  a very surprising result. It means that Weyl transformations are much less trivial than previously thought. It questions, in particular, the full equivalence of Einstein frame and Jordan frame. 
\par
Of course it is possible to take the less rigid point of view that the counterterms are admissible in spite of them being nonlocal. Once this is taken for granted then one quickly realizes that these counterterms are not suppressed by any mass scale (precisely because of conformal invariance). They could then legitimately also be included in the {\em classical} theory.
 Precisely our aim in the present paper is to follow the consequences of this viewpoint and speculate on some of the properties of a hypothetical conformal theory of gravity to all orders in perturbation theory. 
 \par
 
 Conformal invariance for us is {\em exactly} the same as  Weyl invariance  under
\bea
&&\tilde{g}_{\m\n}\equiv \Omega^2(x) g_{\m\n}(x)\nonumber\\
&&\tilde{\psi}=\Omega^{-\l_\psi}\psi
\eea 

Where $\l_\psi$ is by definition the conformal weight of the matter field $\psi$.
To be specific , we will be interested in the four-dimensional action of  CDG, that is.
\begin{align}
S_{CDG}=-\int d^{n}x ~\sqrt{|g|}~\left(-{1\over 12} \phi^2 R - \frac{1}{2}\D_{\m}\f\D^{\m}\f + {g\over 4 !}~ \f^4\right)
\end{align}
which is classically Weyl invariant in a somewhat tautological way, provided the conformal weight of the  field $\f$ is $\l_\f=1$.  This graviscalar field  $\f$ is none other than the Weyl parameter promoted to the range of a physical field. This is sometimes called a {\em St\"uckelberg field} or else a {\em compensating field}. For a mathematician this is simply a {\em group averaging} of sorts.
\par
Conformal physics is not very intuitive in that it does not single out any scale. For example, we all are used to the the idea that 
quantum gravity effects should decouple at energies much smaller than Planck mass, $M_p\equiv {1\over \sqrt{16 \pi G}}$, so that they can be safely ignored in particle physics except in exotic circumstances.
\par
This fails to be true in a conformal theory, as has been explicitly shown in \cite{Alvare} (essentially because there is no any preferred scale in them). Conformal theories are also bizarre in other aspects \cite{Georgi,Weinberg}, notably in the absence of the usual concept of {\em particle}.
\par
The existence of a conformal invariant fundamental theory including the gravitational interaction is however one of the Holy Grails of theoretical physics. Such a hypothetical theory would be extremely interesting, were it only as a theoretical model \cite{Fradkin}. We are here still in the initial steps of this quest.
 
\par           
When quantizing the theory in background field gauge, as has been done in \cite{Alvare}, there is a four dimensional Ward identity stemming from conformal invariance, namely

\bea
&&{\cal D}~ W\left[\bg,\bp\right]\equiv \left.{\d\over \d\Omega(x)}~ W\left[\bg,\bp\right]\right|_{\Omega=1}=\nonumber\\
&&= \left(-2 \bg_{\m\n}{\d \over \d \bg_{\m\n}}+\bp{\d \over \d\bp} \right)W[\bg,\bp]=0
\eea
where the on shell free energy $W[\bg,\bp]$ is the logarithm of the on shell partition function given by the functional path integral
\be
W[\bg,\bp]\equiv -\text{log}~Z[\bg,\bp]
\ee
This is precisely the Ward identity we claim to be anomalous to two-loop order.

\section{Implementation of the Ward identity in perturbation theory through counterterms}
It was shown in \cite{Alvare} that modulo wave-function renormalizations --which are physically irrelevant-- the one-loop UV 
divergence of Conformal Dilaton Gravity is the appropriate Weyl transform of the corresponding one-loop UV divergence of General Relativity found
by in \cite{'tHooftV}, and thus it vanishes. The appropriate Weyl transformation in question reads
\be 
\bg_{\mu\nu}\rightarrow \frac{1}{12 M^2_p}\bp^2\bg_{\mu\nu}
\ee
Furnished with this result and the fact that the classical action of Conformal Dilaton Gravity is obtained from that of General Relativity by applying the previous Weyl transformation, one is led to assume that modulo wave-function renormalizations the two-loop UV divergence of Conformal Dilaton Gravity can be obtained from 
the two-loop counterterm~\cite{Goroff} of General Relativity by applying to the latter the Weyl transformation we have just mentioned. Hence, the resulting two-loop counterterm --see~\cite{Alvare}, for further details--
reads 
\be
W^{L=2}_\infty[\bp,\bg]={1\over n-4}\frac{12}{(4\pi)^4}\frac{209}{2880}\int \sqrt{|g|}~d^n x
~\bp^{-2}~W^{(n)}
_6[\bg],
\ee
in the $n$-dimensional space of Dimensional Regularization. $W^{(n)}_6[\bg]$, which has conformal weight $\lambda=6$, is defined in terms of the
Weyl tensor $W^{(n)\,\a_1\a_2\a_2\a_4}$ in $n$ dimensions as follows
\be
W^{(4)}_6[\bg]=W^{(n)\,\a_1\a_2\a_2\a_4}W^{(n)}{\,\a_3\a_4\a_5\a_6}W^{(n)\,\a_5\a_6}_{\phantom{(n)\,\a_5\a_6\,}\a_1\a_2}
\ee
The previous results make it natural to speculate that the  L-loop divergence in Conformal Dilaton Gravity will be of the form
\be
W^L_\infty[\bp,\bg]={1\over n-4}\int \sqrt{|g|}~d^n x
~\bp^{-2L+2}~\sum_j  g_j P^j_{L+1}[\bg]
\ee
The constants $g_j$ are unknown but calculable coefficients, and $P^j_{(L+1)}\, j=1\ldots N$ stand for the set of purely gravitational terms with conformal weight $\l=2L+2$ (like the trace of the product of $L+1$ Weyl tensors). These terms have mass dimension $2L+2$, so that the full integrand is dimensionless.
An example is the scalar made out of $L+1$ Weyl tensors. The complete set of conformal tensors  is not explicitly known, but this fact is not essential in our argument. 

\par
Let us recall that ${\cal D}$ be given by
\be
{\cal D} = -2 
\bg_{\m\n}{\d \over \d \bg_{\m\n}}+  \bp{\d \over \d\bp}
\ee
Then, in the $n$ dimensional space of  dimensional regularization
\begin{align}
{\cal D}& \left(d(vol)~\bp^{-2L+2}~\sum_j  g_j P^j_{L+1}\right) =\nonumber\\&-(n-4)\left(d(vol)~\bp^{-2L+2}~\sum_j  g_j P^j_{L+1}[\bg]\right)
\end{align}
This means that the integrand of $W^L_\infty[\bp,\bg]$ is an evanescent operator which yields a putative anomaly
\be
{\cal A}^L[\bp,\bg]\equiv -\int d(vol)~ \bp^{-2L+2}\sum g_j P^j_{L+1}[\bg]
\ee
This anomaly-to-be can actually be cancelled by a {\em finite} counterterm
\be
\Delta W^L[\bp,\bg]=\int d(vol)~\bp^{- 2L+2}~\text{log}~\bp\sum g_j P^j_{L+1}[\bg]
\ee
(just because ${\cal D} ~\text{log}~\bp=1$). It is to be remarked that  the integrand of the above expression is again dimensionless, so that there is no room for any dimensionful coupling constant in front.
\par
The full modified action  of CDG will be of the form
\be
W_{CDG}[\bp,\bg]=S_{CDG}[\bp,\bg]+\sum_{L=1}^\infty \left(W_R^L[\bp,\bg]+\Delta W^L[\bp,\bg]\right)
\ee

This modified action obeys
\be
{\cal D} W_{CDG}[\bp,\bg]=0
\ee
so it qualifies as a conformally invariant one.
\par
The finite part of this action (which we propose to reconsider as a classical action of sorts) reads
\begin{align}
W^{\text{class}}_{CDG}[\bp,\bg]&\equiv -\int d(vol)\bigg[~-{1\over 12} \bp^2 \bR - \frac{1}{2}\D_{\m}\bp\D^{\m}\bp +\lambda \f^4+\nonumber\\&~~~\sum_{L=1}^\infty \phi^{- (2L-2)}~\text{log}~\bp\sum_j g_j P^j_{L+1}[\bg_{\m\n}]\bigg]
\end{align}
and does obey instead
\be
{\cal D} W^{\text{class}}_{CDG}[\bp,\bg]=\sum_L {\cal A}^L[\bp,\bg]
\ee

\section{Classical effects of the finite quantum counterterms.}
The counterterms needed to cancel the putative anomalies enjoy two main properties. First of all, they are finite. Besides, and more importantly, there is no {\em small} coupling constant (id est, no $\kappa^2$, because there is no $\kappa$ in the original lagrangian) in front of them; there is no reason why they should be negligible compared with the classical Lagrangian. This justifies a consideration of those terms already at the classical level. The modified scalar EM reads (suppressing hats on the fields from now on)

\begin{align}
&\Box \phi-\dfrac{1}{6}~ R~\phi+4\l\phi^3-\nonumber\\&-\sum_{L=1}^\infty\left(1-(2L-2) \log\phi\right)\phi^{-2L+1}~\sum_j g_jP_{L+1}^j[\bg_{\m\n}] =0
\end{align}
Under a Weyl rescaling the conformal wave operator $\Box_c\equiv \Box -\dfrac{1}{6}~ R~$ behaves as
\be
\Box_c\rightarrow \Omega^{-3}~\Box_c ~\Omega
\ee
It is then possible to write the Weyl transform of the scalar EM
\begin{align}
&\Omega^{-3}\left(\Box -\dfrac{1}{6} R\right)\phi+4\Omega^{-3}\l\phi^3-\nonumber\\&-\Omega^{-3}\sum_{L=1}^\infty\left(1-(2L-2) \log~{\phi\over \Omega}\right)\phi^{-2L+1}~\sum_j g_j P_{L+1}^j[\bg_{\m\n}]=0
\end{align}

\subsection{A one-dimensional toy model}

As the general form of the gravitational terms is not known, a one-dimensional toy model that captures some features of the general setting can be studied. In order to do that, let us assume that $\sum_j g_j P_{L+1}^j[\bg_{\m\n}]\equiv C$ is just a constant independent of $L$.
Then the sums over the loop order can be exactly done
\be
\sum_{L=1}^\infty \phi^{-2L+1}={\phi\over \phi^2-1}
\ee
\be
\sum_{L=1}^\infty (2L-2) \phi^{-2L+1}={2 \phi\over (\phi^2-1)^2}
\ee
which are convergent only when
\be
\left|{1\over \phi}\right| < 1
\ee
and are extended to the whole complex plane by analytic continuation. In that sense, $\phi=0$  is still a solution. Specific analysis are necessary in other situations.

In this case the one-dimensional model then would read
\begin{align}
&{d^2 f(x)\over dx^2}- A f(x)+ B f(x)^3-\nonumber\\&- C~\left({f(x)\over  f(x)^2-1}-{2 f(x)\over (f(x)^2-1)^2}\text{log}~f(x)\right)=0
\end{align}
Consider, besides, $A,B$ as arbitrary constants (where $A$ is proportional to the Ricci scalar and B to the self-coupling $g$).

As the non-invariance of the theory comes from the logarithm, the (toy version of the) conformal case can be recovered in the case $C=0$. In this case the equation can be solved easily by transforming $f(x)\to \lambda f(x)$, then it reads

\be \lambda{d^2 f(x)\over dx^2}- A \lambda f(x)+ \lambda^3 B f(x)^3=0\label{toy}
\ee

Setting $\lambda^2 B=2$ (which can be done as long as $B\neq 0$) and dividing by $\lambda$ we find now

\be{d^2 f(x)\over dx^2}- A  f(x)+   2f(x)^3=0
\ee

finally, calling $m=2-A$ the previous equation now reads

\be{d^2 f(x)\over dx^2}- (2-m)  f(x)+   2f(x)^3=0
\ee

which is solved by the Jacobi elliptic function $\text{dn}(x|m)$.

\begin{minipage}{0.5\textwidth}
	\centering
	\includegraphics[scale=0.25]{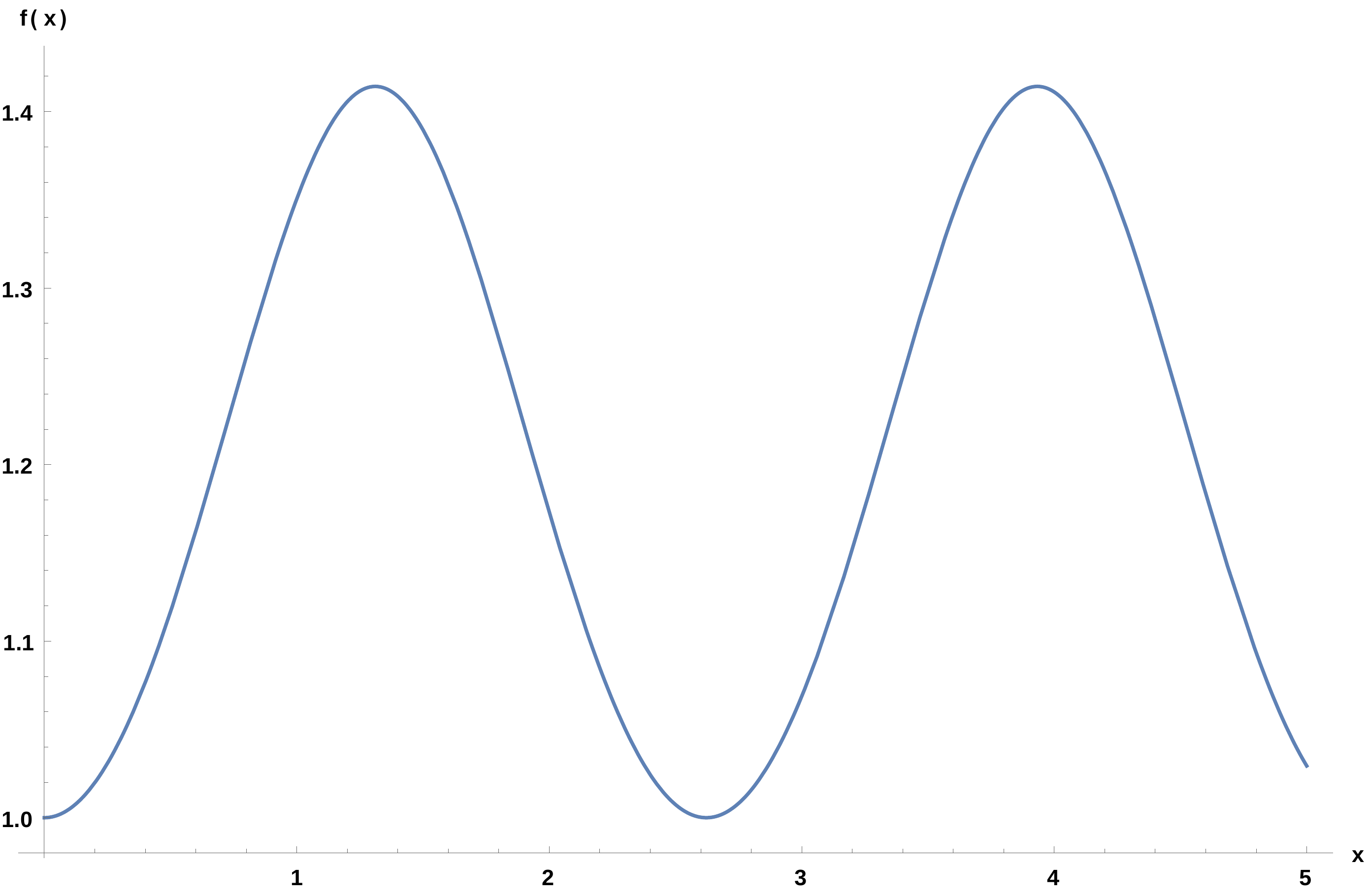}
	\captionof{figure}{}
	
\end{minipage}

This function is defined in terms of the elliptic integral
\be u=\int_0^\phi\dfrac{d\theta}{\sqrt{1-m\sin^2\theta}}\ee

where m is called the \textit{parameter}. Then the function we are dealing with is defined as $\text{sn}~ (u|m)=\sin \phi$.

In order to see explicitly the properties of this solution, it is useful to define the quarter-periods $K$ and $iK'$, in the following way.

Let $m_1$  (the \textit{complementary parameter}) be such that $m+m_1=1$, then

\begin{align}
K&=\int_0^{\tfrac{\pi}{2}}\dfrac{d\theta}{\sqrt{1-m\sin^2\theta}}=\dfrac{\pi}{2}\sum_{n=0}^{\infty}\left(\dfrac{(2n)!}{2^{2n}(n!)^2}\right)^2 m^n\\
iK'&=i\int_0^{\tfrac{\pi}{2}}\dfrac{d\theta}{\sqrt{1-m_1\sin^2\theta}}=i\dfrac{\pi}{2}\sum_{n=0}^{\infty}\left(\dfrac{(2n)!}{2^{2n}(n!)^2}\right)^2 (1-m)^n
\end{align}

In terms of these, the function has periods $2K$, $4K+4iK'$ and $4iK'$. As long as we have a real solution here, we are only concerned by $2K$ which can be expressed (to the lowest order) in terms of A as 

\be 2K=\dfrac{3\pi}{4}-\dfrac{\pi}{8}A\ee

thus its period is smaller as bigger is the curvature.

We consider now the non-conformal case (i.e $C\neq 0$) given by \eqref{toy}. The first thing to notice is that due to the presence of the logarithm $f(x)=0$ is no longer a solution. In fact, this is expected as we are working in the broken phase ($\phi\neq 0$). This is the main difference between the conformal and non-conformal case which is also expected in a more complicated model.

Concerning the constants,
neither the sign of $A$ or (i.e. the curvature) or $C$ change the shape of the function. Depending on the initial conditions there are two possibilities
\begin{enumerate}
	\item If $C>0$ there is an interval where the solution  is again  periodic but and looks as the conformal case. (Although the ODE diverges for $f(x)=1$, numeric integration yields finite answer even in this case)
	
	\begin{minipage}{0.4\textwidth}
		\centering
		\includegraphics[scale=0.4]{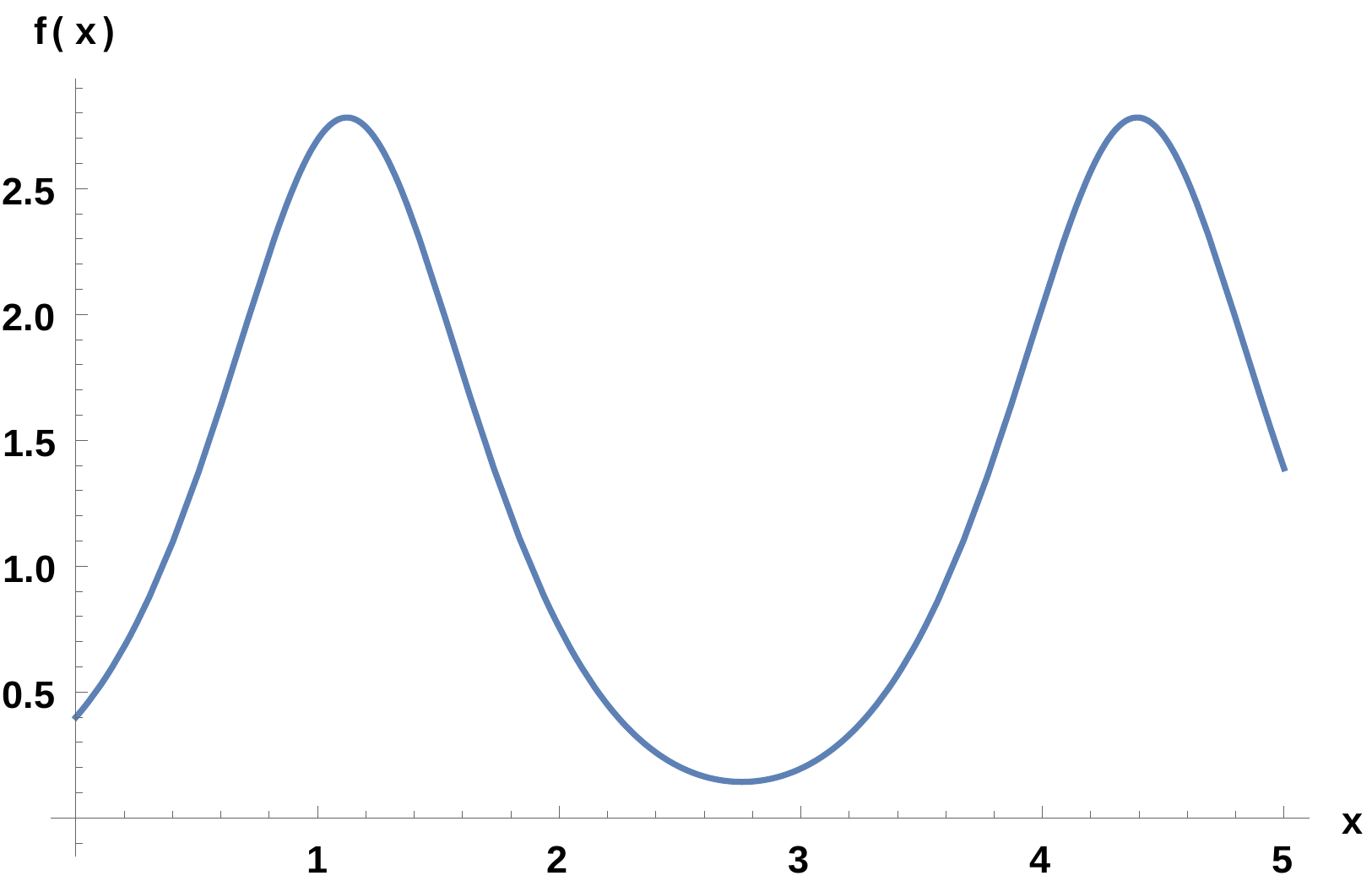}
		\captionof{figure}{}
	\end{minipage}

	\item In any other case \ref{fig:3} the solution does not oscillate and goes to zero.
	
	\begin{minipage}{0.4\textwidth}
		\centering
		\includegraphics[scale=0.4]{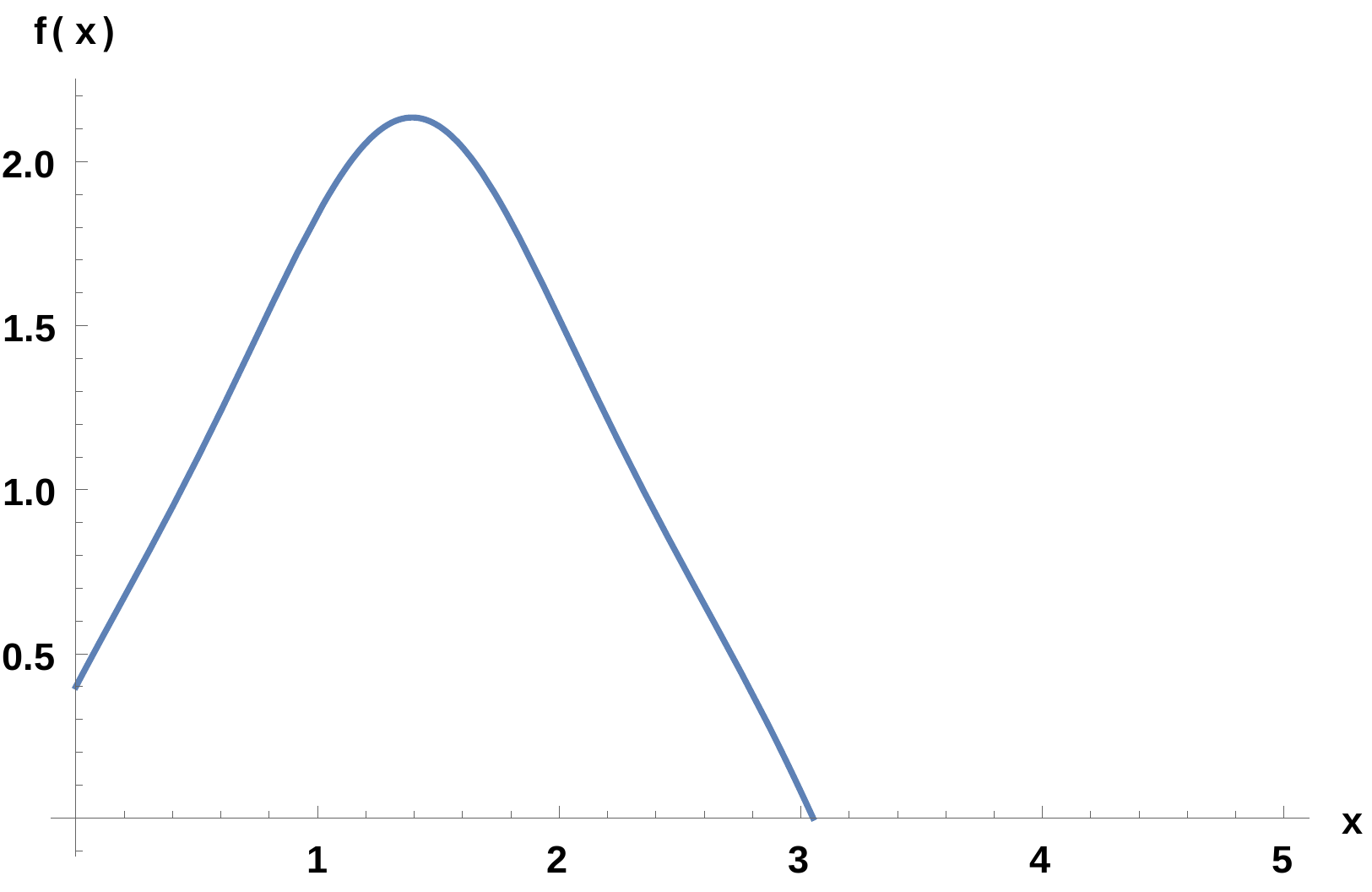}
		\captionof{figure}{$f'(0)>0$}
		\label{fig:3}
	\end{minipage}	
	\begin{minipage}{0.4\textwidth}
		\centering
		\includegraphics[scale=0.4]{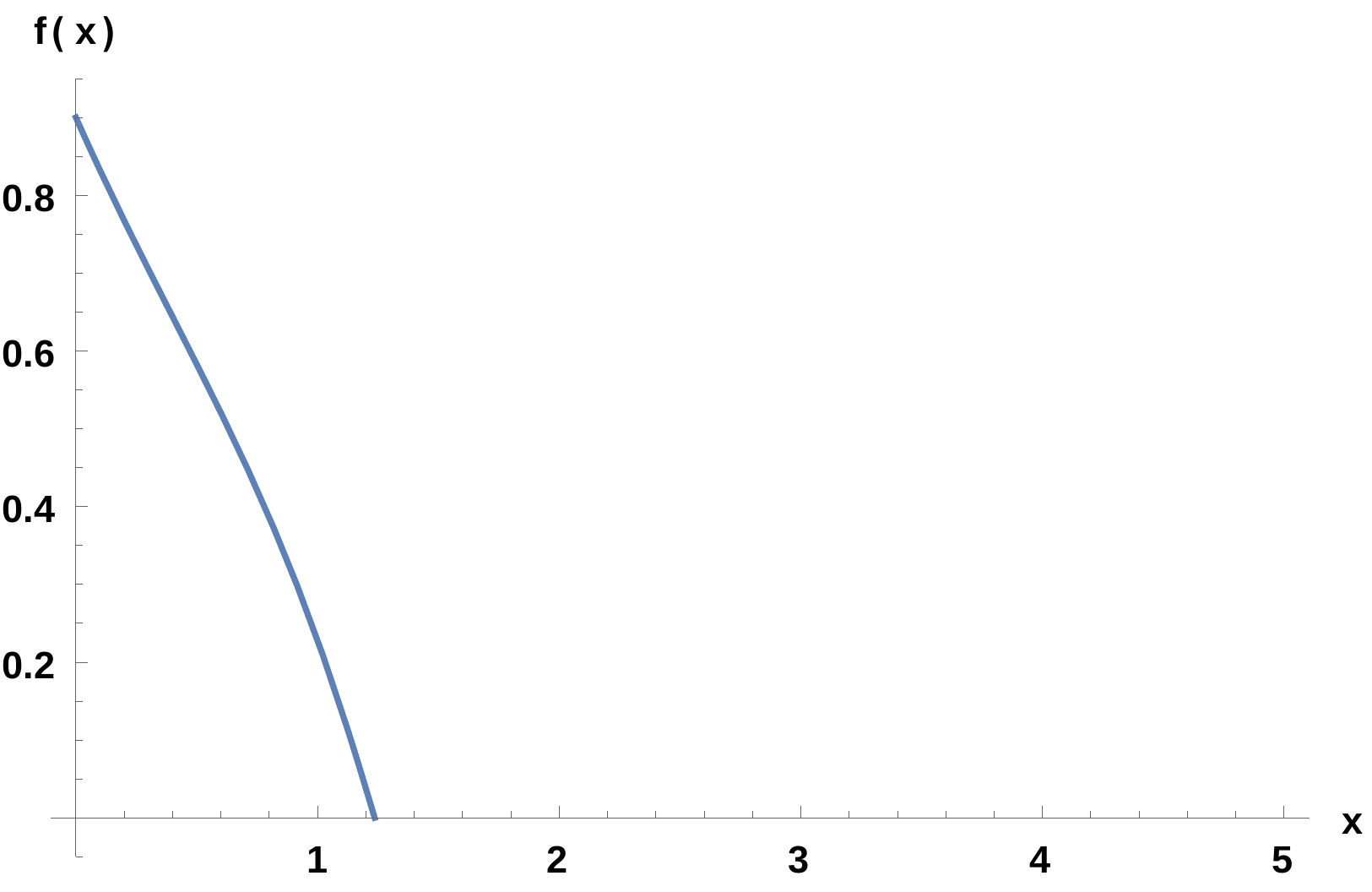}
		\captionof{figure}{$f'(0)<0$}
		\label{4}
	\end{minipage}

\end{enumerate}

To summarize,  the inclusion of the non-conformal part does not change greatly the solution. Depending on the initial values it can be periodic as in the conformal one, or  go to zero and vanish.

There is another difference between both theories. In the conformal case, in addition to the trivial solution, the constant values $f(x)=\pm\sqrt{\dfrac{A}{B}}$ are solutions of the equations of motion. However, this does not happens in the non-conformal model, where there are not any constant solutions.

Although our toy model is one-dimensional, it is likely that it embodies some of the characteristics of the full fledged four-dimensional situation.

\section{The symmetric phase of CDG}
It is not obvious how the symmetric phase of CDG (which corresponds in the background field language to $\bp=0$) should be understood. The first problem is that there is no propagator to damp the gravitational fluctuations in the loop approximation. This has been emphasized in particular by 't Hooft \cite{Hooft}. Nevertheless there are some  observations that can be made on general grounds. The full partition function can be written as

\bea
&Z[\bg_{\m\n}]\equiv\int {\cal D }\phi{\cal D} h_{\m\n}e^{-\int d(vol)\bigg[ \phi\left(\bn^2-{1\over 6}\overline{R}\right)\phi+~O(\phi^2 h,h^3)+O(\phi^2 h^2)\bigg]}
\eea
Please note that the integral over the ghosts and auxiliary fields are implicitly included in the measure ${\cal D} h_{\a\b}~{\cal D}\phi$. Also note that the gauge-fixing conditions that suit our analysis should contain no monomial linear in the fields $h_{\mu\nu}$ and $\phi$. This way the gauge-fixing terms will contain three or more quantum fields.
\par
The one loop contribution only involves the quantum fluctuations of the dilaton. Its divergent part can be easily computed:
\begin{align}
&Z^{(L=1)}_\infty[\bg_{\m\n}]=\text{exp}~\left(-{1 \over 16 \pi^2}{1\over n-4}\int d(vol)\left({1\over 180}\left( \overline{R}_{\m\n\r\s}^2-\nonumber\right.\right.\right.\\&\qquad\qquad\qquad-\left.\left.\left.\overline{R}_{\m\n}^2+\overline{\Box}~ \overline{R}\right)\right)\right)=\nonumber\\
&=\text{exp}~\left(-{1 \over 16 \pi^2}{1\over n-4}\int d(vol){1\over 180}\left({3\over 2} \overline{W}_4-{1\over 2}\overline{E}_4+\overline{\Box}~ \overline{R}\right)\right)
\end{align}
We have represented by $W_4$ the square of Weyl tensor
\be
W_4\equiv R_{\m\n\r\s}^2 - 2 R_{\m\n}^2+{1\over 3} R^2
\ee
and by $E_4$ the four-dimensional Pfaffian (which upon integration yields Euler's characteristic up to a constant)
\be
E_4\equiv R_{\m\n\r\s}^2-4 R_{\m\n}^2+ R^2
\ee
We are not able to perform the loop integrals over the gravitational fluctuations because there is no propagator for the gravitational field. Assuming (and this is an explicit hypothesis) that this integral makes sense (for example by discretizing the system) then the partition function can be defined as given by the expression 
\be
 Z_\infty=\sum_{L=1}^\infty Z_\infty^L[\bg_{\m\n}]
\ee
which is the sum of all higher loop divergent pieces $Z_\infty^L[\bg_{\m\n}]$ each of which  is a  conformal invariant functional of $\bg_{\m\n}$. There is only one of those in four dimensions, namely
\be
Z_\infty[\bg_{\m\n}]\equiv e^{-g \int d(vol) \overline{W}_4}
\ee
All loop contributions are of the same form, so that we can represent by $g$ the coefficient of the whole sum. This procedure is formal in more that one way; there is no reason in particular to expect the loop expansion to converge or even to be asymptotic.
\par
\section{Conclusions.}

The general form of the the finite counterterms which is necessary to insure cancellation of the Weyl anomaly to every order in perturbation theory has been determined --under a sensible assumption,
using only conformal invariance. They involve logarithms of the physical scalar, so that they are not local terms {\em sensu stricto}.
 \par

 We found it interesting to examine the most broad minded hypothesis in which they are indeed acceptable counterterms. Then two facts immediately come to our attention. First of all, and in spite of their being loop effects (and so carrying powers of $\hbar$, so to speak)  those finite counterterms do not have any inverse power of any mass scale in front of them (precisely because of conformal invariance) and then they are not negligible in the low energy deep infrared limit. This might be identified in some sense with the {\em classical limit}.
\par
It is then of interest to consider the classical effects of those terms. The most important such is that the status of the  trivial conformal invariant  solution
\be
\phi=0
\ee 
changes slightly. This solution is the only compatible with the symmetric phase of conformal symmetry.
\par
When the space-time  is of Petrov type O (that is, Weyl flat) then the symmetric configuration  is still a solution. When the Weyl tensor does not vanish, then the analysis is more involved. 

Consider the oversimplified situation in which  the purely gravitational contribution is L-independent
\be
\sum_j g_j P_{L+1}^j [\bg_{\m\n}]\equiv G(x)
\ee
Then the scalar equation of motion reduces to
\bea
&&\left(\Box-{1\over 6}~R+{g\over 6} \phi^2\right)~\phi+\nonumber\\
&&+\left( 2 {\phi\over (\phi^2-1)^2}\log~\phi-{\phi\over \phi^2-1}\right)~G(x)=0
\eea
 In that sense, $\phi=0$  is still a solution. Specific analysis are necessary in other situations.
In order get an idea how to do that a one-dimensional toy model has been studied.
 A comparison was made between results without taking into account the counterterms (this a conformally invariant situation which can be exactly solved in terms of Jacobian elliptic functions) and results including terms in our toy model that mimick the said counterterms.
 \par
 It is to be stressed that in spite of the above, those are {\em not} the classical equations of motion to be used in the context of the background field gauge technique to express physical results on shell, Those correspond to the $\hbar=0$ sector only; that is without including the corrections studied in the present paper. The fact that the solution corresponding to the symmetric phase is not always admissible in the present setting  has to be interpreted as the fact that in those cases the counterterms are necessarily singular in the symmetric phase.
 \par
  Finally we conjecture that the form of the symmetric phase of conformal dilaton gravity ought to be proportional to the Weyl squared theory.

\section*{Acknowledgments}
 We are  grateful for helpful discussions with Manuel Asorey and Mario Herrero-Valea. We  also acknowledge useful email exchange with Michael Duff as well as discussions with Roman Jackiw and So Young Pi. Part of this work was done while E.A. was on the {\em Aspen Institute of Physics} and on  the Lawrence Berkeley Laboratory. We have been partially supported by the European Union FP7  ITN INVISIBLES (Marie Curie Actions, PITN- GA-2011- 289442)and (HPRN-CT-200-00148) as well as by FPA2012-31880 (Spain), FPA2014-54154-P, COST action MP1405 (Quantum Structure of Spacetime) and S2009ESP-1473 (CA Madrid).  The authors acknowledge the support of the Spanish MINECO {\em Centro de Excelencia Severo Ochoa} Programme under grant  SEV-2012-0249. 
\appendix
\section{Conformal invariants}

Let us summarize here some known facts about conformal (Weyl) invariants.
The Schouten tensor is defined as
\be
A_{\a\b}\equiv {1\over n-2}\left(R_{\a\b}-{1\over 2(n-1)} R g_{\a\b}\right)
\ee

It is invariant under rigid Weyl rescaling, that is, it transforms under $\Omega\equiv e^{\s}$ as
\be
\tilde{A}_{\a\b}=A_{\a\b}-\s_{\a\b}-{1\over 2}(\nabla\s)^2 g_{\a\b}
\ee
The Weyl tensor reads
\be
W_{\a\b\m\n}\equiv R_{\a\b\m\n}+\left(A_{\b\m}g_{\a\n}+A_{\a\n}g_{\b\m}-A_{\b\n}g_{\a\m}-A_{\a\m}g_{\b\n}\right)
\ee
 It transforms as a conformal tensor of weight $\l=-1$:
\be
\tilde{W}_{\a\b\m\n}\equiv e^{2\s}~W_{\a\b\m\n}
\ee
Its square has got scale dimension $\l=2$
\be
\tilde{W}_{\a\b\m\n}~\tilde{W}^{\a\b\m\n} = e^{-4\s}~W_{\a\b\m\n}~W^{\a\b\m\n}
\ee
in such a way that
\be
|g|^{2/n} W^2
\ee
is pointwise invariant (but behaves as a true scalar in four dimensions only).
\par
The Weyl tensor  vanishes identically in low dimension $n=2$ and $n=3$. An space with $n\geq 4$ is conformally flat iff $W=0$.
\par
The Cotton tensor reads
\be
C_{\a\b\g}\equiv \nabla_\a A_{\b\g}-\nabla_\b A_{\a\g}
\ee
is a conformal invariant of scaling dimension $\l=0$ in $n=3$ dimensions (and only there). 
\par
It is traceless in any dimension.
\begin{align}
g^{\m\n} C_{\a\m\n}&=\nabla_\a A-g^{\m\n}\nabla_\m A_{\a\n}\nonumber\\&={1\over 2(n-1)}\nabla_\a R-{1\over 2(n-1)}\nabla_\a R=0
\end{align}

\par
 The Bach tensor reads
 \be
 B_{\m\n}\equiv \nabla^\r C_{\r\m\n}+A^{\a\b}W_{\a\m\n\b}=\nabla^\a\nabla_\d W_{\a\m\n\d}-{1\over 2} R^{\a\d} W_{\a\m\n\d}
 \ee
 (this fact stems from the second Bianchi identity).
 
 The Bach tensor is transverse
 \be
 \nabla_\b B^{\a\b}=0
 \ee
 
 and it inherits its tracelessness from the same property for Weyl and Cotton tensors
 \be
 g^{\m\n}B_{\m\n}=0
 \ee
 It is a conformal invariant of scaling dimension $\l=1$ in four dimensions only.

The variation of the four-dimensional Weyl-squared action  yields  precisely the Bach tensor
\be
\d \int |W|^2 d(vol)=\int B_{\m\n}\d g^{\m\n} d(vol)
\ee
It is of course well known that that there is an extension of the Laplacian 
\be
\Box_c\equiv \Box-{n-2\over 4(n-1)}~R
\ee
that is such that
 \be
 \tilde{\Box}_c=\Omega^{-{n+2\over 2}}~\Box_c~\Omega^{n-2\over 2}
 \ee
\par
On the other hand, the operator (which is a total derivative)
\be
\Box_2\equiv \sqrt{-g}~\Box
\ee
transforms as
\be
\tilde{\Box_2}=\pd_\m\left(\Omega^{-2} g^{\m\n}\pd_\n\right)
\ee

The quartic  Paneitz operator in arbitrary dimension 
\be
Q(g)\equiv \Box^2+\nabla_\n\left(-{4\over n-2}R^{\m\n}+{n^2-4n+8\over 2(n-1)(n-2)}~R~g^{\m\n}\right)\pd_\m
\ee
is conformal invariant in the same sense as the conformal Laplacian is; that is, under 
\be
\tilde{g}_{\a\b}\equiv \Omega^2 g_{\a\b}
\ee
transforms as 
\be
\tilde{Q}=\Omega^{-{n+4\over 2}}~Q~\Omega^{n-4\over 2}
\ee
In four dimensions this gives 
\be
\Delta_P\equiv \sqrt{-g}~\left(\Delta^2+2\nabla_\m\left(R^{\m\n}-{1\over 3}R~g^{\m\n}\right)\nabla_\n\right)
\ee

 The Fefferman-Graham (FG) {\em obstruction tensor} $O_{\m\n}$ \cite{Fefferman} is a trace-free symmetric two-tensor which has got scaling dimension
 \be
 \l={n-2\over 2}
 \ee
 and is divergenceless
 \be
 \nabla^\l O_{\m\l}=0
 \ee
  and vanishes for conformally Einstein metrics. It is the Dirichlet obstruction to the existence of a formal power series solution for the ambient metric associated to a given conformal structure. For example, the equation
  \be
  R_{\a\b}[g_+]+n g^+_{\m\n}=O\left(x^{n-1}~\text{log}~x\right)
\ee
admits a solution of the form
\be
g^+_{\m\n}={1\over x^2}\left(dx^2+g^x_{\m\n}\right)
\ee
where
\be
g^x_{\m\n}=h^x_{\m\n}+r^x_{\m\n} x^n \text{log}~x
\ee
Then
\be
n~c_n~r^0_{\m\n}=O_{\m\n}
\ee
where
\be
c_n\equiv {2^{n-2}\left(n/2 -1\right)!^2\over n-2}
\ee

\par
 In higher dimension $n\geq 6$ Graham and Hirachi \cite{GrahamH} have shown that the Weyl tensor and the FG obstruction are the basic building blocks of conformal invariants.
 Although explicit formulas are not known, it can be shown that
 \begin{align}
 O_{\m\n}&=\Delta^{n/2 -2}\left(\Delta P_{\m\n}-\nabla_\n\nabla_\m P_\l^\l\right)+\text{lots}=\nonumber\\
 &={1\over 3-n}\Delta^{n/2 -2}\nabla^\r\nabla^\s~W_{\s\m\n\r}+\text{lots}
 \end{align}

 There is also an analogue of the four-dimensional Weyl action \cite{Branson}, namely, the Q-curvature \cite{Chang}, which is not a point wise conformal invariant, but yields nevertheless a conformal invariant under integration on a compact manifold; in fact \cite{Alexakis} $\int Q$ is a combination of the Euler characteristic and the integral of a point wise conformal invariant.
 \par
 It is related to the conformally invariant n-th power of the Laplacian $P_n$ and under Weyl $\tilde{g}=e^\s~g$,
 \be
 e^{n \s\over 2}~\tilde{Q}=Q+P_n~{\s\over 2}
 \ee
 Given the fact that $P_n$ is self-adjoint and annihilated constants, the preceding result follows.
 \par
Consider an asymptotic expansion of the volume
\be
\text{Vol}_{g_+}\left(\e<x<\e_0\right)=c_0~ \e^{-n} + c_2~\e^{-n+2} + \ldots + c_{n-2}~\e^{-2} + \text{L}~\text{log}{1\over \e}+O(1)
\ee
The logarithmic term is related to the integral of the Q-curvature
\be
\int Q dv=(-1)^{n/2}n(n-2)c_n~ L
\ee

 \be
 \d\int_M Q\sqrt{|g|}~d^n~ x=(-1)^{n\over 2}~{n-2\over 2}~\int_M \sqrt{|g|}~d^n x~O_{\m\n}~\d g^{\m\n}
 \ee
 

\end{document}